\documentclass[journal]{IEEEtran}
\usepackage{amsmath,amsfonts}
\usepackage{algorithmic}
\usepackage{algorithm}
\usepackage{array}
\usepackage[caption=false,font=normalsize,labelfont=sf,textfont=sf]{subfig}
\usepackage{textcomp}
\usepackage{stfloats}
\usepackage{url}
\usepackage{verbatim}
\usepackage{graphicx}
\usepackage{cite}
\usepackage{booktabs}

\usepackage[colorlinks=true,
            linkcolor=blue,   
            citecolor=blue,  
            urlcolor=black]{hyperref}
\hypersetup{pdfborder={0 0 0}}
\usepackage[normalem]{ulem} %
\usepackage{soul} %

\usepackage{xcolor} %
\usepackage{orcidlink}          
\usepackage{tikz}
\renewcommand{\arraystretch}{1.0}

\begin{document}


\title{Enhanced-BLE: A Hybrid BLE-ESB Framework for Dynamically Reconfigurable and Energy-Efficient 2.4 GHz IoT Communication}

\author{Ziyao Zhou\orcidlink{0009-0007-0671-9947},
        Chen Shen\orcidlink{0009-0008-4221-2534},
        Tiancheng Cao\orcidlink{0000-0002-7259-5192},
        Hen-Wei Huang\orcidlink{0000-0003-1921-8897}%
\thanks{This work was supported by the Nanyang Assistant Professorship, the MOE Tier 1 grants RG71/24 and RG26/25, and the MTC MedTech Programmatic Fund M24N9b0125. \textit{(Corresponding author: Chen Shen.)}}%
\thanks{Ziyao Zhou, Chen Shen, Tiancheng Cao and Hen-Wei Huang are with the School of Electrical and Electronic Engineering, Nanyang Technological University, Singapore (e-mail: ZHOU0557@e.ntu.edu.sg; shenchen@ntu.edu.sg; tiancheng.cao@ntu.edu.sg; henwei.huang@ntu.edu.sg). Hen-Wei Huang is also with the Lee Kong Chian School of Medicine, Nanyang Technological University, Singapore.
}

}

\markboth{IEEE Internet of Things Journal
,~Vol.~XX, No.~X, Month~Year}%
{Shell \MakeLowercase{\textit{et al.}}: A Sample Article Using IEEEtran.cls for IEEE Journals1}


\maketitle
\begin{abstract}
Bluetooth Low Energy (BLE) is the dominant wireless protocol for IoT systems due to its low power and reliable bidirectional communication. However, its connection-oriented architecture introduces fundamental trade-offs in wake-up latency, throughput, and energy efficiency, limiting its suitability for burst-mode and on-demand event-driven sensing applications. In contrast, Enhanced ShockBurst (ESB), a lightweight connectionless protocol operating on the same 2.4 GHz Nordic Semiconductor hardware, enables ultra-fast wake-up and highly efficient data transmission, but lacks the robustness and reliability of BLE for sustained bidirectional communication. This work systematically investigates the complementary strengths and limitations of BLE and ESB on a unified Nordic nRF54L15 platform, and proposes Enhanced-BLE, a hybrid framework that integrates BLE and ESB to extend the capabilities of conventional BLE systems. Experimental benchmarking shows that ESB nearly halves packet transmission time and energy compared with BLE, doubles the achievable forward throughput, and reduces wake-up latency and wake-up energy by nearly twentyfold during intermittent operation. However, ESB reverse transmission may suffer packet loss, whereas BLE maintains reliable bidirectional communication.
To address these limitations, Enhanced-BLE employs adaptive radio scheduling and coexistence-aware connection management to combine ESB-based high-throughput transmission with BLE-based reliable reverse communication. The framework enables protocol handover from BLE to ESB within approximately 18 ms and restores BLE operation within 49 ms from standby mode. Experimental results demonstrate that Enhanced-BLE achieves approximately twofold higher forward throughput than BLE while significantly reducing warm-up latency. These results demonstrate a practical and hardware-compatible strategy to overcome the intrinsic limitations of BLE, enabling low-latency, high-throughput, energy-efficient, and reliable wireless communication for next-generation IoT and on-demand sensing systems.

\end{abstract}

\section{Introduction}

The rapid growth of Internet-of-Things (IoT) systems has intensified the demand for wireless communication solutions that can adapt to diverse traffic patterns, including intermittent event-driven transmission, high-throughput data bursts, and reliable bidirectional feedback \cite{Ferdowsy2021Traffic, Senadhira2023UAVAssisted}. In many emerging applications, devices must remain in low-power standby states for extended periods \cite{Raut2025EnergyEfficient}, respond rapidly after wake-up\cite{Radfar2020Battery, Deshpande2023EnergyEfficient}, transmit burst data efficiently when needed\cite{Whitcomb2025Bursty,Zhao2019Fast}, and maintain reverse communication for control or feedback \cite{Huang2019Optimal}. Under these conditions, communication efficiency is determined not only by steady-state throughput and power consumption, but also by transient wake-up behavior, protocol overhead, and bidirectional link reliability.

Among existing low-power wireless technologies, Bluetooth Low Energy (BLE) has become a widely adopted standard due to its interoperability, scalability, and mature ecosystem \cite{Gomez2012Overview, Koulouras2025Evolution, Tosi2017Performance}. However, it exhibits several limitations when applied to complex and dynamic IoT scenarios. BLE connectionless advertising can support wake-up communication \cite{Gautam2024LowPower}, but transmitted packets are prone to loss due to the absence of acknowledgment (ACK) and retransmission mechanisms \cite{10106496}. Such unreliability is unacceptable for critical data transmission. On the other hand, BLE's connection-oriented communication can preserve data reliability, but it introduces non-negligible delay during device discovery and connection establishment, often referred to as warm-up time, which can significantly delay responses \cite{cho2014analysis}. Additionally, the maximum physical layer (PHY) supported by BLE is 2 Mbps \cite{Badihi2020On}, which constrains its capability to efficiently accommodate dynamic or data-intensive communication demands. These limitations are especially problematic for IoT applications where timely access to critical data and efficient burst transmission are required.

In contrast, Nordic's proprietary Enhanced ShockBurst (ESB) protocol \cite{veilleux2018esb} could be an alternative communication mechanism tailored for power-sensitive IoT and event-driven applications. Originally developed for low-latency, packet-based wireless communication \cite{11337444}, ESB enables rapid data exchange without the need for complex connection procedures, thereby significantly reducing warm-up time and improving responsiveness. Its lightweight protocol design minimizes communication overhead while incorporating built-in ACK and retransmission mechanisms to ensure robust and reliable one-direction data delivery. Additionally, ESB supports a faster 4 Mbps PHY than BLE, making it particularly suitable for burst data retrieval scenarios that demand intensive communication. However, ESB reverse transmission cannot be actively initiated by the receiver and may suffer from packet loss.

These contrasting characteristics highlight a fundamental trade-off in IoT wireless communication: standardized protocols provide interoperability and scalability at the cost of efficiency, while lightweight protocols offer performance advantages but lack ecosystem support \cite{8528458}. Despite the complementary characteristics of BLE and ESB, a systematic comparison between the two protocols remains insufficiently explored.

Motivated by this gap, we present a comprehensive experimental benchmark of BLE and ESB on a unified hardware platform, evaluating key performance metrics including packet-level time and energy, system power consumption, throughput, wake-up overhead, and bidirectional capability. Based on these insights, we further propose a hybrid Enhanced-BLE framework, which enables seamless integration of BLE and ESB within a single device using the Multi-Protocol Service Layer (MPSL) \cite{nordic_mpsl_readme}. By leveraging the shared hardware compatibility of the two protocols, the proposed approach enables dynamic, lightweight ESB operation and standardized BLE communication, effectively combining their complementary advantages.

This proposed Enhanced-BLE framework provides a practical pathway to address the limitations of both protocols. In event-driven scenarios with frequent sleep--wake transitions, ESB can maintain data transmission before BLE service discovery is completed. When higher forward bandwidth is required, ESB can be used to extend the throughput capability beyond BLE-only communication. Conversely, when ESB lacks reliable reverse data transmission, the BLE reverse link can be used to ensure stable feedback communication. In addition, the framework supports low-latency handover between BLE and ESB, enabling the system to adapt to changing traffic requirements. Overall, Enhanced-BLE offers a promising solution for next-generation IoT systems requiring adaptive, high-performance, and energy-efficient wireless communication.

The main contributions of this work are summarized as follows:

\begin{itemize}
    \item A systematic and fair experimental benchmark is established to compare BLE and ESB on an identical hardware platform, covering packet-level time consumption and energy, continuous throughput--power scaling, wake-up overhead, and bidirectional communication.

    \item The key performance advantages and limitations of BLE and ESB are quantitatively characterized. The results show that ESB provides higher throughput, lower packet energy, and substantially reduced wake-up overhead compared with BLE, while its reverse-link reliability, interoperability, and security remain limited compared to BLE.

    \item A hybrid Enhanced-BLE framework is proposed and experimentally validated to combine the complementary advantages of both protocols. By integrating ESB-assisted wake-up, ESB-forward transmission, BLE-based reverse communication, and separate transmission power (TXP)/PHY control, the framework improves the responsiveness, throughput, flexibility, and energy efficiency of IoT communication.
\end{itemize}

\section{Methodology for Comparing BLE and ESB Communication}

\begin{figure}[t]
    \centering
    \includegraphics[width=0.9\linewidth]{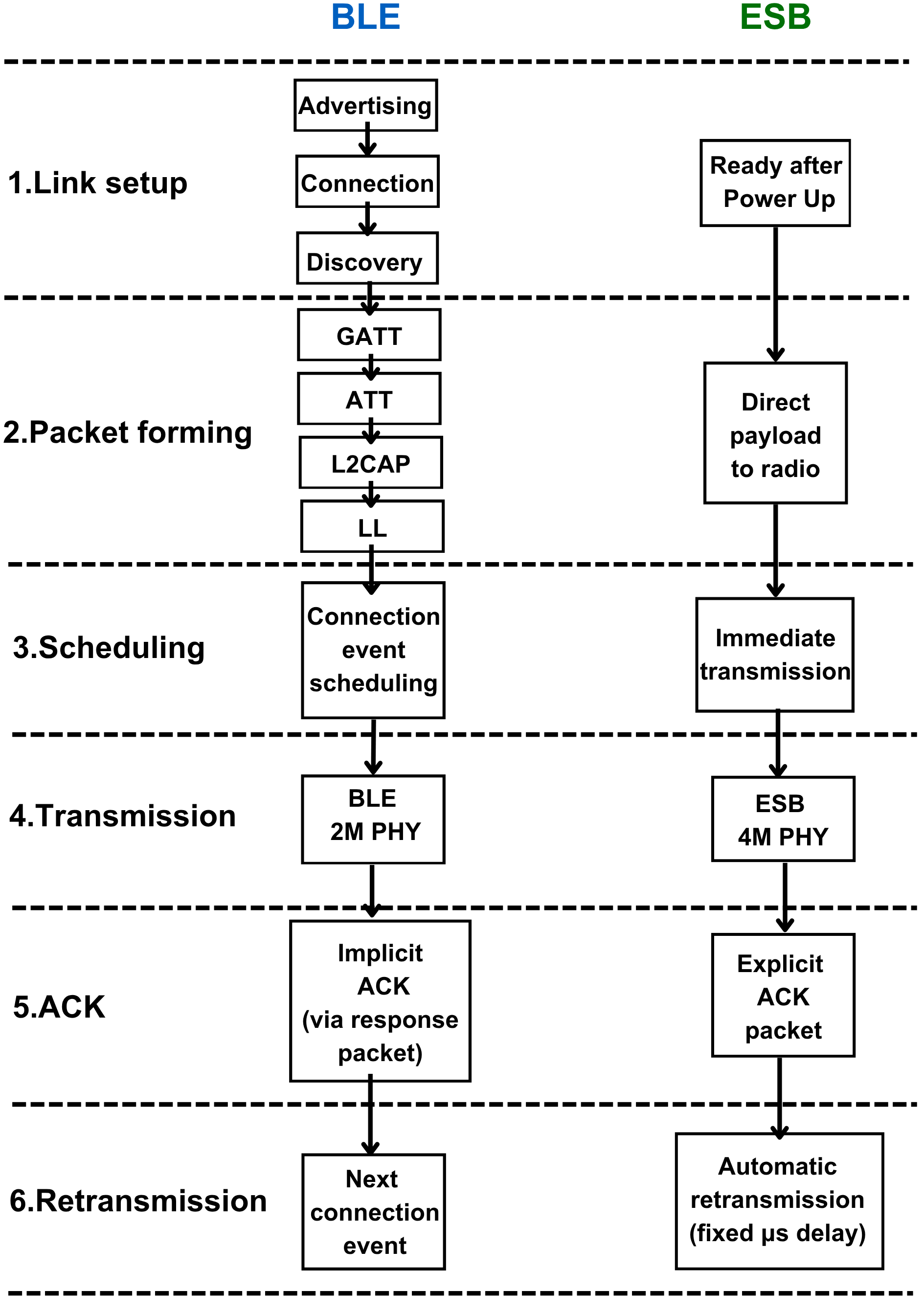}
    \caption{Comparison of single transmission procedures between BLE and ESB. BLE relies on connection-oriented scheduling, whereas ESB enables connectionless execution with lightweight packet handling.}
    \label{fig:ble_esb_flow}
\end{figure}

Both BLE and ESB can run on the same Nordic wireless microcontroller unit (MCU) with identical hardware. To enable a fair and interpretable comparison, their differences should therefore be first examined at the software level for a single transmission. Specifically, each protocol adopts a distinct procedure for data processing, scheduling, and transmission, which directly determines the measured performance metrics.

As illustrated in Fig. \ref{fig:ble_esb_flow}, BLE and ESB follow different procedures for a single transmission. BLE requires a link establishment process, including advertising, scan response, connection, and service discovery, before data transmission, whereas ESB can directly enter the transmission state.

During packet formation, BLE processes data through multiple protocol layers (GATT, ATT, L2CAP, and link layer), while ESB directly forwards the payload to the radio. For scheduling, BLE transmissions are constrained by connection events, whereas ESB supports immediate transmission without predefined time slots.

At the physical layer, BLE operates up to 2M PHY, while ESB operates up to 4M PHY, resulting in different transmission speeds. In terms of acknowledgement, BLE relies on implicit ACK within connection events, whereas ESB uses explicit ACK packets. For retransmission, BLE may schedule retransmissions in subsequent connection events \cite{11222604}, while ESB performs automatic retransmission with a fixed microsecond-level delay, by default, 600 µs.

\subsection{Comparison Between BLE and ESB for Single-Packet Transmission}

To evaluate the practical single-packet energy and time consumption of BLE and ESB communication, a pair of nRF54L15 development boards was employed as the experimental platform, which could be programmed with either BLE or ESB firmware. Both the transmitter and the receiver were placed in close proximity under line-of-sight conditions in the air. For BLE, the connection interval was set at 7.5 ms. The advertising interval (20 ms) and scanning interval (2.5 ms) were also set to their minimum values to minimize device discovery time and ensure fast link establishment. In contrast, ESB does not require explicit configuration of a connection. In addition, both communication schemes were configured to operate at their highest available PHY. BLE employs the 2M PHY, whereas ESB utilizes a proprietary 4M PHY. Both protocols were evaluated independently with a TXP of 8 dBm.

For a fair comparison, both protocols were configured to transmit 244-byte packets, which correspond to the maximum payload supported by BLE and remain below the ESB maximum payload limit of 252 bytes.

The transmitter was connected to a Power Profiler Kit II (PPK), which sampled the supply current at 100,000 samples/s to capture the instantaneous current consumption \cite{nordicsemi_ppk2_userguide}. For each BLE and ESB packet transmission event, the corresponding time and energy consumption were measured using the PPK and subsequently compared between the communication protocols.

\subsection{Comparison Between BLE and ESB Under Continuous-Packet Transmission}

To investigate how the previously observed differences in single-packet transmission accumulate during continuous communication, the effective data throughput was adjusted by varying the inter-packet time delay. This approach enabled the operation to be swept from a 0 kbps idle condition to the maximum achievable throughput of each protocol. In this experiment, BLE was configured with its maximum payload of 244 bytes, while ESB utilized its maximum payload of 252 bytes. During the experiment, the transmitter and receiver were positioned in close proximity in the air to characterize the power consumption associated with continuous data transmission.

In addition, understanding the relationship between Received Signal Strength Indicator (RSSI) and throughput is critical for characterizing the achievable maximum throughput of the two protocols under different attenuation conditions and application scenarios \cite{11222604}. To experimentally investigate this relationship, the receiver was fixed while the transmitter was gradually moved away, enabling RSSI to be swept over a range from approximately -90 dBm to -30 dBm, and the corresponding throughput was measured.

\subsection{Comparison Between BLE and ESB Under Sleep-Wake Operation}

Long-term IoT monitoring systems often adopt an event-driven, on-demand sensing strategy, in which the device remains in an ultra-low-power sleep state for extended periods and wakes only briefly to acquire and transmit data. This duty-cycled operation effectively reduces the average power consumption and prolongs battery lifetime 
\cite{Traverso2026Miniaturized}. However, BLE is less suited to such an operation because each wake-up event requires a relatively long warm-up period to re-establish connection, introducing additional delay and energy overhead. In contrast, ESB operates in a connectionless manner and therefore may enable faster wake-up, making it more suitable for on-demand IoT applications.

To compare the performance of BLE and ESB in on-demand monitoring scenarios, both communication protocols were configured to wake up every 10 s, perform one packet transmission, and then return to sleep. A General Purpose Input/Output (GPIO) pin was used as a state indicator to mark key events, including completion of system initialization, BLE connection establishment, and completion of the first packet transmission.  The PPK was used to record the energy consumption during the duty cycle on the transmitter side, allowing the warm-up period to be analyzed in terms of both time and energy consumption. 

\subsection{Comparison Between BLE and ESB in Bidirectional Communication}

Despite its potential advantages in throughput and energy efficiency, ESB theoretically lacks a symmetric bidirectional communication capability to BLE. BLE supports independent data transfer in both directions within each connection event \cite{Gomez2012Overview}. In contrast, ESB adopts a master-initiated transaction model, where all communication is triggered by the transmitter. The receiver can only return data by embedding it within ACK packets, rendering the reverse link entirely dependent on forward-link activity.

To systematically characterize the similarities and differences between the two protocols in bidirectional communication, the packet transmission rate of one node is controlled, while the peer node transmits in the reverse direction at its maximum achievable throughput under the current condition. Therefore, the obtained results represent the upper-bound performance of the system under each bidirectional configuration.

For BLE, the packet transmission rate at the peripheral device was adjusted to regulate the throughput observed at the central device, while the payload size was fixed at 244 bytes, and the BLE connection interval was maintained at 7.5 ms. For ESB, the packet transmission rate of the transmitter is controlled to regulate the throughput at the receiver, using a payload size of 252 bytes. The receiver then responds by sending ACK packets at its maximum capability under the current configuration. To investigate the impact of reverse-link payload size, the ACK packet size is varied among 2 bytes, 132 bytes, and 252 bytes.

To quantify the reliability of ACK-based reverse transmission in ESB, the reverse data transfer was initiated by the transmitter, while the reverse ACK payload carried an incrementing sequence number. The transmitter recorded the received ACK sequence numbers, identified missing indices, and calculated the ACK packet loss rate accordingly. RSSI was varied by changing the separation distance between the two nodes and was logged at 1 s intervals to investigate the relationship between ACK loss and RSSI.

\section{Methodology for Integrating BLE-ESB Communication}

BLE–ESB coexistence is implemented using the Nordic MPSL, which provides an efficient radio scheduling framework. Instead of allowing concurrent access, both protocols share a single transceiver through time-division multiplexing, where BLE maintains priority for connection events, and ESB operates within dedicated timeslots scheduled by MPSL.

\begin{figure}[t]
    \centering
    \includegraphics[width=0.9\linewidth]{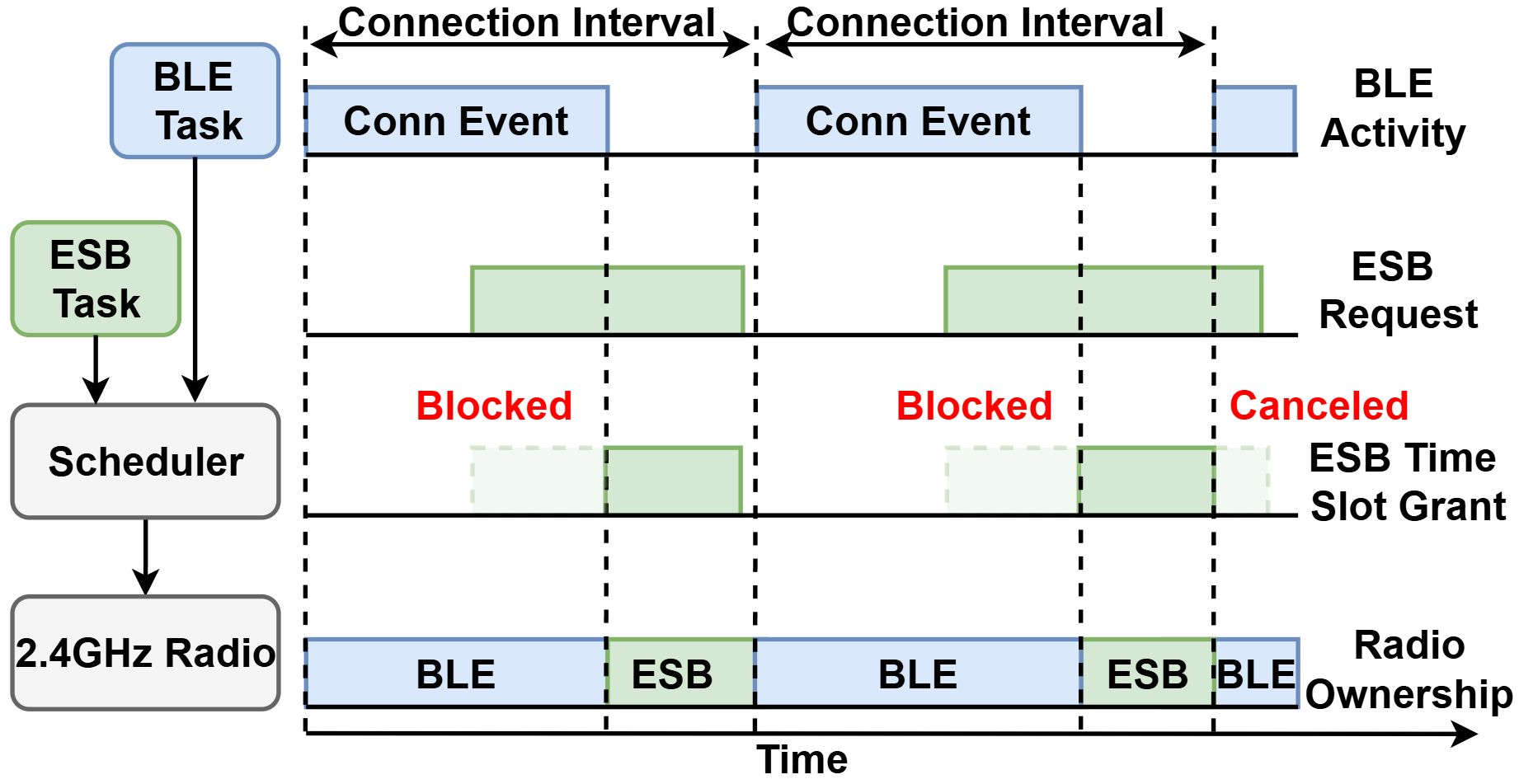}
 \caption{MPSL-based radio scheduling between BLE and ESB. BLE connection events are prioritized, while ESB requests are granted only when the radio is available.}
    \label{fig:mpsl_ble_esb_scheduler}
\end{figure}

During operation, BLE continuously manages connection intervals and creates corresponding connection events within each interval for data transmission. In parallel, ESB requests radio access according to its own communication requirements. The MPSL schedules and arbitrates radio access between the two protocols \cite{Zhou2026Multiprotocol}. MPSL prioritizes BLE to maintain the connection interval and event. Therefore, when BLE is transmitting data during a connection event, and ESB also requests radio access, the ESB request is blocked by MPSL. Conversely, when ESB is transmitting data, and a BLE connection event occurs, BLE preempts the radio resource, and the ongoing ESB transmission is canceled. The overall scheduling process is illustrated in Fig. \ref{fig:mpsl_ble_esb_scheduler}.

Because radio access, mode switching, and timing are coordinated by MPSL at a higher protocol management layer above both BLE and ESB stacks, the coexistence mechanism avoids packet collision and minimizes timing uncertainty introduced by software-level scheduling. This ensures that BLE connection stability is preserved while enabling opportunistic ESB communication without persistent interference.

The proposed framework was experimentally validated using three nRF54L15 development kits. One device implements the Enhanced-BLE firmware and operates concurrently as the BLE peripheral transmitter and the ESB transmitter. The remaining two devices serve as independent receivers: one acts as a BLE central device to receive BLE packets, while the other operates as an ESB receiver. Each receiver computes and reports real-time performance metrics, including throughput and RSSI for both BLE and ESB links. This setup enables direct, side-by-side evaluation of the two protocols under identical operating conditions.

\subsection{Enhanced-BLE Coexistence Configuration}

In the proposed system, BLE is treated as the primary protocol whose communication performance must be preserved during coexistence. Consequently, its configurable parameters directly influence the coexistence behavior. Among these parameters, the connection interval is particularly critical \cite{7964929}, as it determines the temporal spacing for data exchange between the central and peripheral devices.

To characterize its impact on coexistence performance, the BLE connection interval was swept across three values: 7.5 ms, 100 ms, and 1000 ms. During the evaluation, the BLE and ESB transmitters were operated concurrently using their respective maximum payload sizes, with 244-byte packets for BLE and 252-byte packets for ESB in transmitter mode, consistent with the previous configuration. 

To further investigate the coexistence trade-off between BLE and ESB, the BLE throughput was dynamically controlled, while ESB was operated at its maximum transmission rate adaptively. For each operating point, the simultaneously achievable BLE and ESB throughput and the corresponding total system power consumption were measured. Therefore, the obtained results represent the upper-bound performance of the system under each coexistence configuration.

\subsection{Enhanced-BLE Handover Mechanism}
When Enhanced-BLE is used, the requirements and allocation of BLE and ESB can be dynamically adjusted, which inevitably introduces additional switching overhead. The handover scenarios can be mainly categorized into three types. First, BLE and ESB transmit concurrently, while their bandwidth requirements vary over time, requiring handover between the two links. Second, only one protocol is active at a given time, while the other remains in standby mode and is activated through handover when needed. Third, only one protocol is active at a given time, while the other is completely powered off and subsequently re-enabled through handover.

When BLE and ESB operate concurrently, the handover latency during dynamic radio-resource allocation is calculated as follows. The BLE handover latency is defined as the time interval between issuing the command to modify the BLE throughput and the next BLE transmission pattern. Similarly, the ESB handover latency is defined as the time interval between issuing the command to modify the ESB throughput and the transmission of the next ESB transmission pattern. Neither protocol provides an interface to directly determine when the new pattern is physically transmitted over the air; only the time at which the packet is written into the protocol stack can be obtained. Therefore, in this study, the handover latency is extracted from the interval between commands and power variation measured during the switching process.

Similarly, for the standby and complete shutdown scenarios in which only one protocol is active at a given time, the switching latencies from ESB to BLE and from BLE to ESB are obtained by analyzing the measured power waveforms.

\subsection{Enhanced-BLE Wake-Up Mechanism}
Since BLE requires connection establishment before communication can begin, a communication gap exists between system wake-up and BLE link availability. In contrast, ESB operates in a connectionless manner and can transmit data immediately after initialization. Therefore, ESB is introduced to bridge this wake-up gap in the proposed framework. After waking up, the system first transmits data through ESB while BLE advertising, connection establishment, and service discovery proceed in the background. Once the BLE connections are fully configured, ESB transmission is terminated, and communication is handed over to BLE. This strategy maintains communication continuity during the wake-up process and avoids the long period during which BLE-only systems are unable to transmit application data.

The advertising interval of the peripheral device was changed from the minimum value of 20 ms to 100 ms to emulate a typical IoT device under normal operating conditions \cite{9672120, 8795488}. The scanning interval of the central device was kept unchanged.

During this wake-up experiment, two GPIO pins on the development board were configured as protocol status indicators corresponding to BLE and ESB activity, respectively. Four key protocol stages were identified during the wake-up process: ESB start, BLE connect, BLE start, and ESB stop after BLE takeover. At each stage transition, the corresponding GPIO signal was pulled low for 50 ms as a timing marker. Both GPIO signals were connected to the logic analysis module of the PPK, enabling synchronized measurement of system power consumption and protocol operating states.
\subsection{Enhanced-BLE Bidirectional Communication}

Due to the lightweight nature of the ESB protocol, reverse communication from the ESB receiver to transmitter can only be achieved by embedding the payload within the reverse ACK packet. However, the ACK packet itself cannot be independently initiated and does not support acknowledgment or retransmission mechanisms, which may result in packet loss during reverse transmission. In contrast, BLE employs acknowledgment and flow control mechanisms for bidirectional communication, enabling reliable and stable data exchange.

Furthermore, the BLE 2M PHY operates at a lower rate than the ESB 4M PHY. As a result, ESB could achieve theoretically twice the throughput of BLE under comparable conditions, or equivalently, lower power consumption for the same throughput requirement.

Based on these complementary characteristics, we propose using the Enhanced-BLE strategy in which ESB is utilized for forward data transmission to maximize throughput and energy efficiency, while BLE is employed for reverse communication to ensure reliable and lossless data delivery. To evaluate this approach, the trade-off relationship between forward and reverse throughput in conventional BLE communication was characterized under an optimal connection interval. Subsequently, the throughput trade-off was evaluated using ESB for forward transmission and BLE for reverse transmission. During the evaluation, each protocol was operated at its maximum achievable transmission rate under the corresponding configuration, and thus the obtained results represent the upper-bound performance of the system.

\subsection{Enhanced-BLE TXP and PHY Control}

To evaluate the protocol-specific configurability of the Enhanced-BLE framework, runtime configuration was performed via the receiver-side interface, allowing BLE and ESB parameters, including TXP and PHY settings, to be separately adjusted during coexistence. These parameters were controlled because TXP and PHY directly affect link quality, communication robustness, and energy consumption under varying wireless conditions.

For TXP characterization, the BLE TXP was swept across +8, -2, and -12 dBm while keeping the ESB TXP fixed at +8 dBm. The procedure was then repeated with ESB TXP varied over the same range, while BLE TXP remained constant. Throughout the experiments, the RSSI of both BLE and ESB links was continuously and synchronously monitored at their respective receiver nodes, enabling direct observation of cross-protocol interference and parameter decoupling.

For PHY characterization, the BLE PHY was dynamically reconfigured from the coded PHY (S8) to the 1M PHY and then to the 2M PHY using a control command transmitted over the link, while ESB transmission remained active. Subsequently, the ESB PHY was reconfigured from the 1M PHY to the 2M PHY and then to the 4M PHY using a separate control command, while BLE transmission remained active. Throughout this process, the throughput of both BLE and ESB links was continuously and synchronously monitored to evaluate the effects of PHY reconfiguration and protocol coexistence.

\section{Results of BLE and ESB Communication Comparison}


\subsection{Comparison Between BLE and ESB for Single-Packet Transmission}

\begin{figure}[t]
\centering
\includegraphics[width=0.9\linewidth]{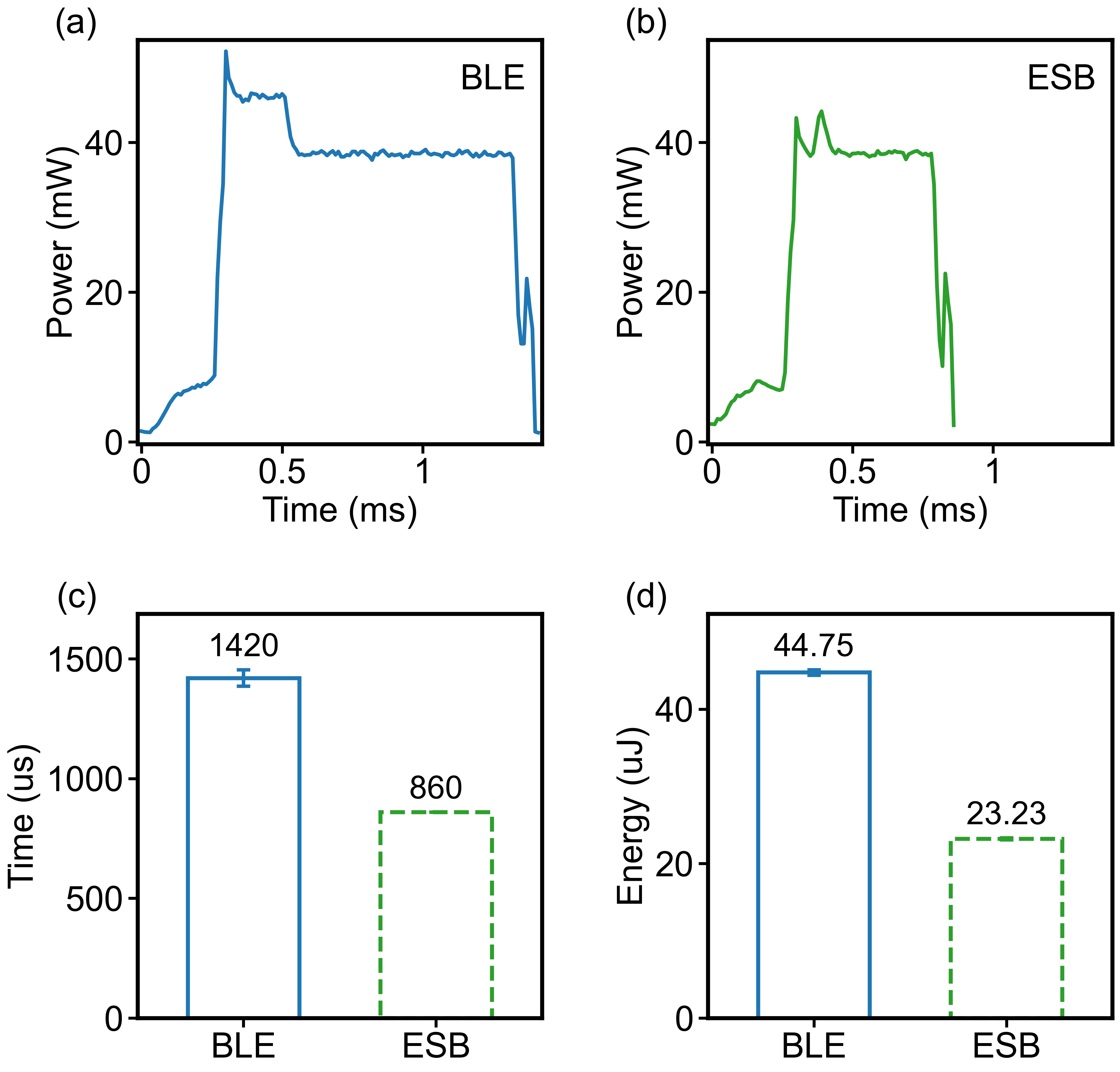}
\caption{Single-packet transmission performance of BLE and ESB. (a) Power consumption profile of BLE during transmission of a 244-byte packet. (b) Power consumption profile of ESB during transmission of a 244-byte packet. (c) Comparison of single-packet transmission time. (d) Comparison of single-packet energy consumption.}
\label{fig:ble_esb_single_packet_characteristics}
\end{figure}

Fig. \ref{fig:ble_esb_single_packet_characteristics} (a) and (b) show the power consumption during the transmission of a single 244-byte packet for two protocols. As observed from the comparison, the overall power profiles of the two protocols exhibit similar shapes, both resembling a trapezoidal pattern. However, a significant difference in transmission duration is observed between the two protocols as summarized in Fig. \ref{fig:ble_esb_single_packet_characteristics} (c), transmitting a single packet using the BLE 2M PHY requires an average of approximately 1420 µs, whereas ESB with the 4M PHY requires only about 860 µs, which is around half that of BLE. Correspondingly, the energy consumption per packet for BLE is nearly twice that of ESB, as illustrated in Fig. \ref{fig:ble_esb_single_packet_characteristics} (d). Specifically, BLE consumes approximately 44.75 µJ per packet, while ESB consumes about 23.23 µJ. These results highlight the clear advantage of the ESB 4M PHY, as transmitting the same data packet requires only about half the time and energy compared with BLE.

\subsection{Comparison Between BLE and ESB Under Continuous-Packet Transmission}

\begin{figure}[t]
    \centering
    \includegraphics[width=0.9\linewidth]{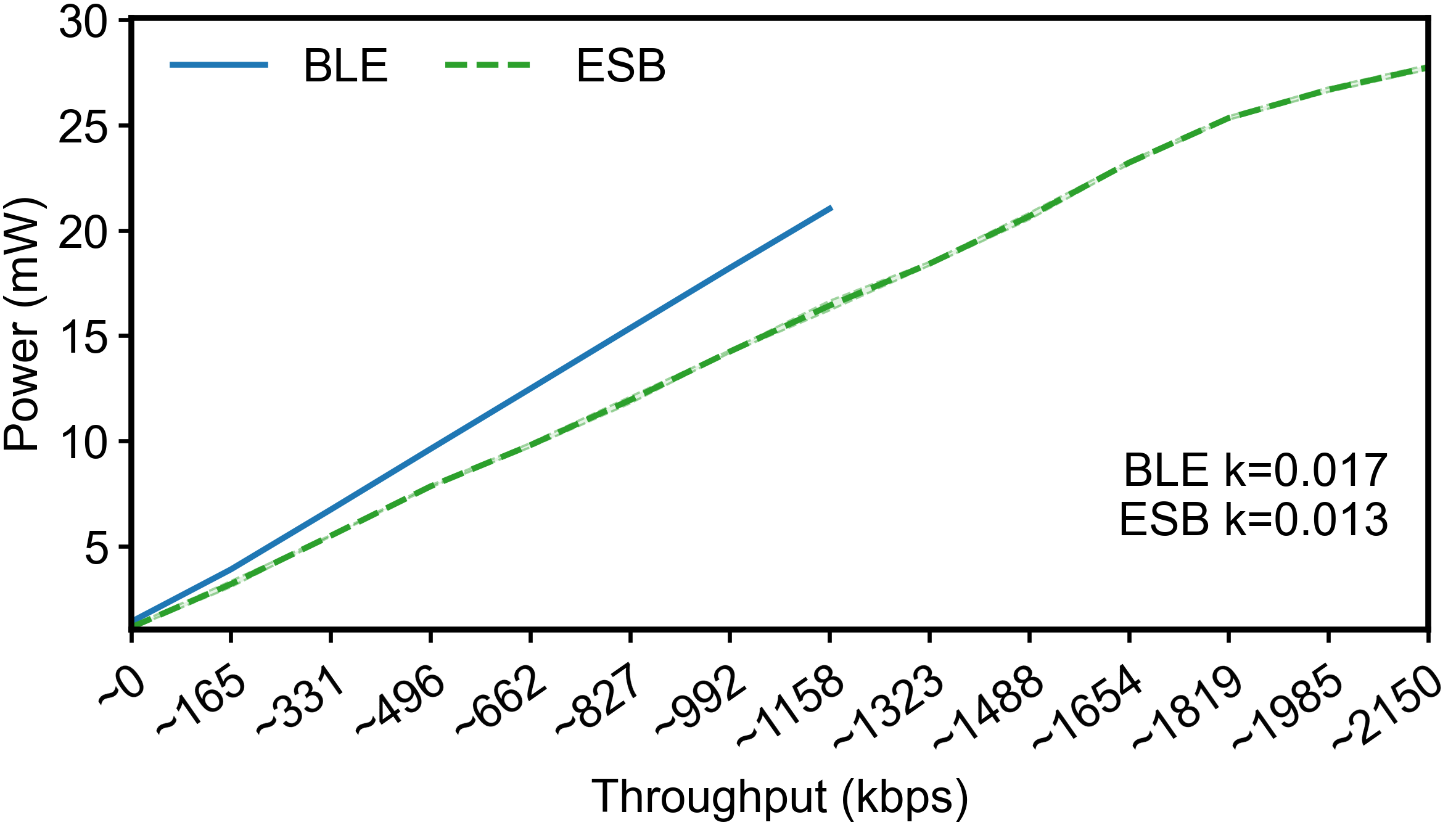}
    \caption{Power consumption versus throughput for BLE and ESB during continuous transmission.}
    \label{fig:power_throughput_ble_esb}
\end{figure}

As the differences observed in single-packet transmission accumulate over time, the performance gap becomes more evident during continuous operation. The standby power consumption and the power consumption during continuous data streaming are summarized in Fig. \ref{fig:power_throughput_ble_esb}. It can be observed that, when no data are transmitted, the standby power consumption of ESB is 0.55 mW, which is lower than that of BLE (0.99 mW). This difference arises because ESB is a more lightweight protocol that does not require maintaining an active connection.


Notably, ESB enables nearly a twofold increase in transmitted data throughput—rising from approximately 1100 kbps for BLE to around 2200 kbps. This improvement is primarily attributed to the higher link capacity enabled by the 4M PHY. The comparative analysis of the two protocols further reveals that the power–throughput scaling slope for BLE is approximately 0.017 mW/kbps, whereas that for ESB is about 0.013 mW/kbps. This indicates that, at an identical throughput level, ESB can reduce power consumption by approximately 25\% compared with BLE, which suggests the potential for longer battery lifetime under communication-dominated operating conditions.

It is worth noting that the approximately 25\% power saving at the same throughput is lower than the approximately 50\% per-packet energy reduction reported earlier. This discrepancy is mainly due to the fixed packet-level overhead in ESB. Specifically, ESB requires a separate ACK packet and a dedicated RX window for each forward transmission, whereas BLE implicitly carries acknowledgment information in the SN/NESN fields of the reverse PDU header. In addition, ESB introduces a fixed 600 µs Auto Retransmit Delay between retransmissions, while BLE can retransmit within the same connection event after a single interval of 150 µs. Because these overheads are defined in absolute time, they are not reduced by increasing the PHY data rate. Therefore, the 4M PHY shortens only the payload on-air time, while the fixed overhead remains unchanged, reducing the effective energy saving from approximately 50\% per packet to approximately 25\% under continuous transmission.

\begin{figure}[t]
\centering

\includegraphics[width=0.9\linewidth]{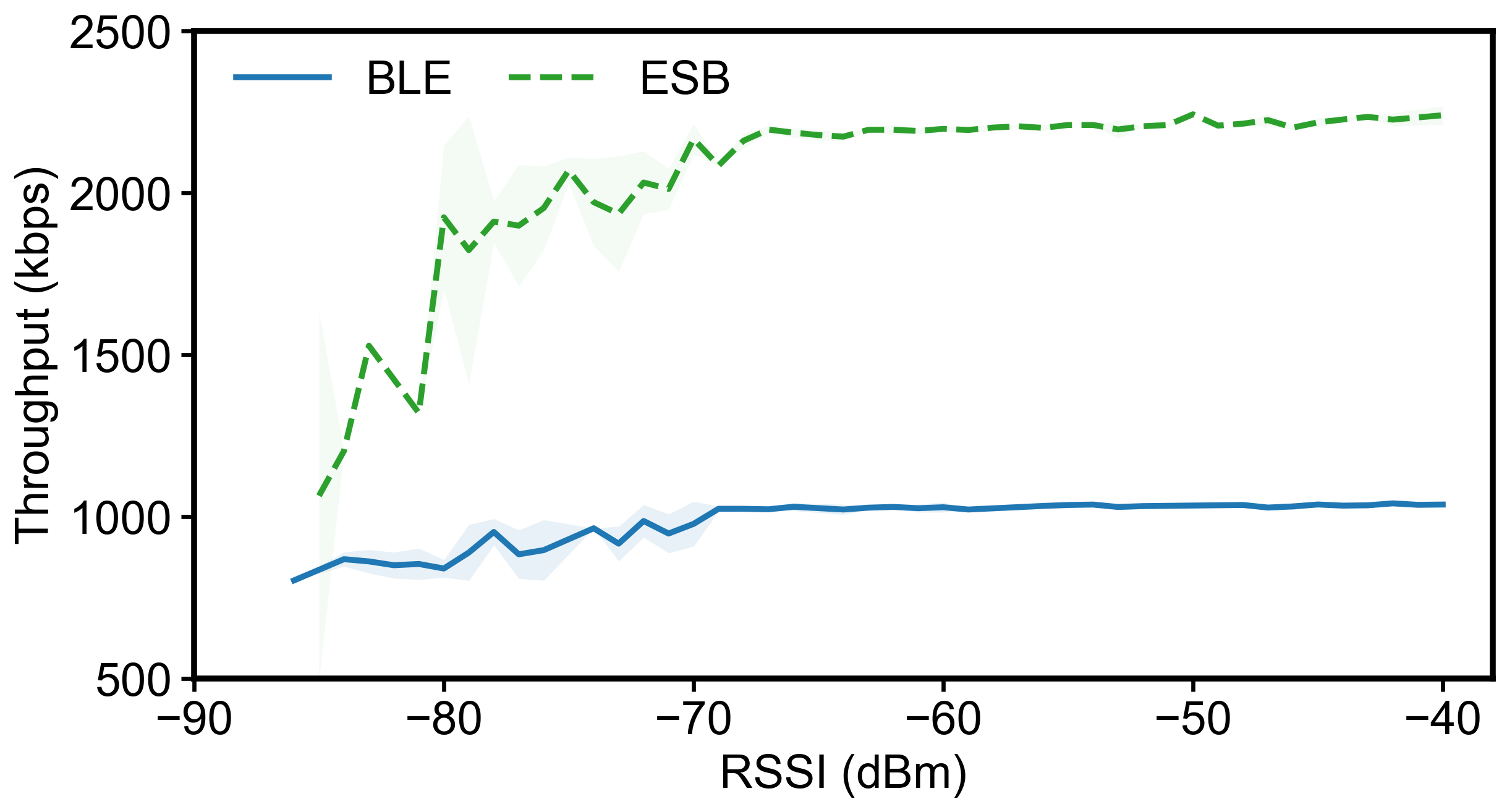}
\caption{Relationship between RSSI and throughput for BLE and ESB.}
\label{fig:rssi_throughput}
\end{figure}

In addition, Fig. \ref{fig:rssi_throughput} illustrates the relationship between RSSI and throughput, which was characterized by artificially adjusting the attenuation level under the same environmental conditions. As shown in the results, a consistent trend can be observed for both protocols: as the RSSI decreases, the achievable throughput also declines. Within the higher RSSI range from -65 dBm to -40 dBm, both protocols are able to maintain near-maximum throughput. 

However, it can be observed that in the lower RSSI range from -85 dBm to -70 dBm, the throughput of ESB decreases much faster than that of BLE. This is because the higher 4M PHY of ESB is more sensitive to signal quality compared with the BLE 2M PHY. As the RSSI decreases and the signal-to-noise ratio degrades, the demodulation performance of the higher PHY deteriorates more rapidly, resulting in a more rapid decline in achievable throughput. Nevertheless, when averaged across this attenuation range, ESB still maintains a higher average throughput than BLE.

\subsection{Comparison Between BLE and ESB Under Sleep-Wake Operation}

\begin{figure}[t]
\centering
\includegraphics[width=0.9\linewidth]{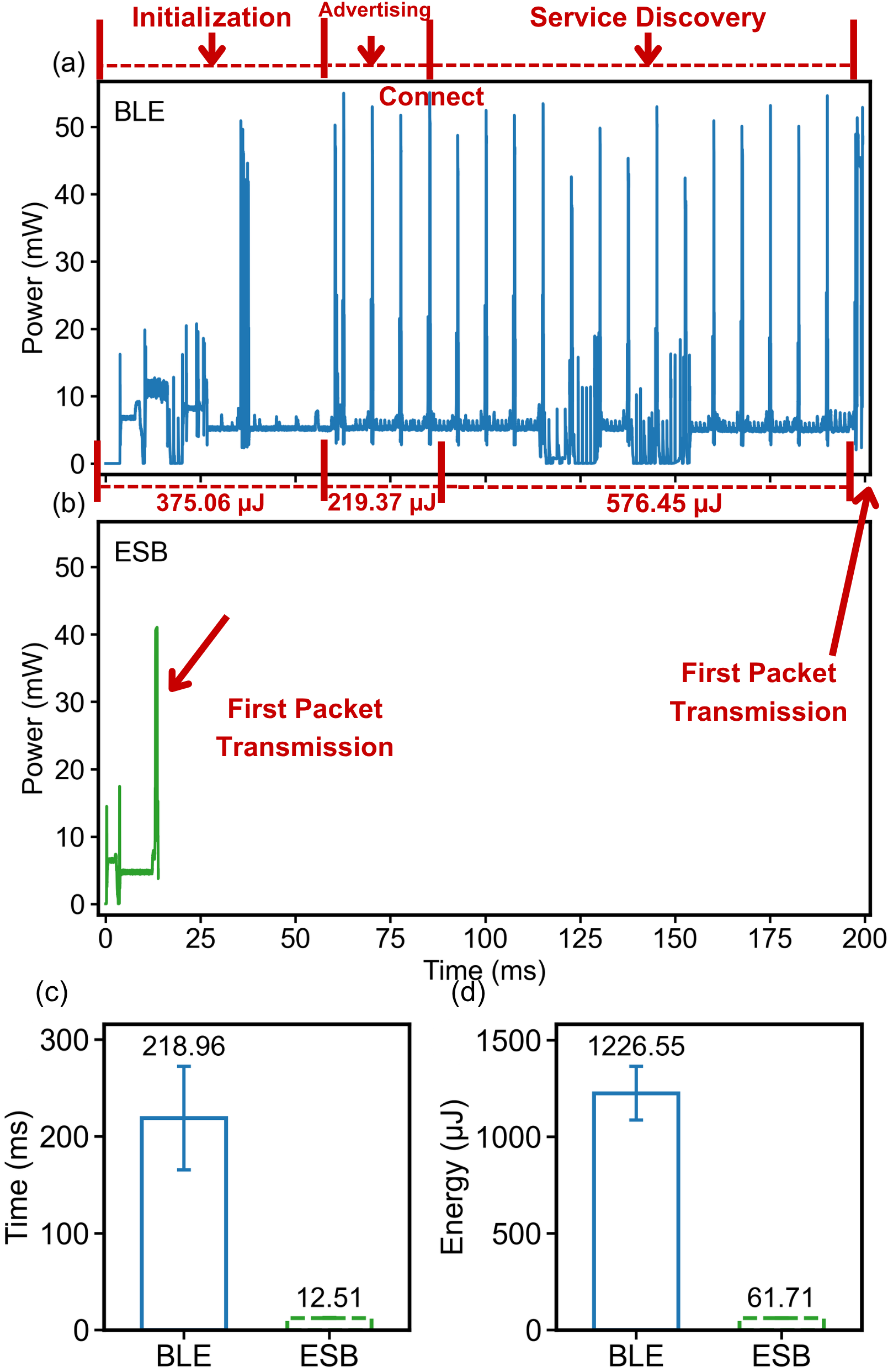}
\caption{Wake-up and first-packet transmission performance of BLE and ESB. (a) Power consumption profile of BLE during wake-up and first-packet transmission. (b) Power consumption profile of ESB during wake-up and first-packet transmission. (c) Comparison of wake-up time. (d) Comparison of energy consumption.}
\label{fig:ble_esb_wakeup_first_packet}
\end{figure}






The comparison between BLE and ESB under sleep-wake communication is shown in Fig. \ref{fig:ble_esb_wakeup_first_packet}. It illustrates the sequence in which each protocol wakes from sleep, initializes the system, transmits one maximum data packet, and subsequently returns to sleep.

Fig. \ref{fig:ble_esb_wakeup_first_packet}(a) shows that, after BLE wakes up and completes the necessary program initialization, it must still undergo advertising, scan response, connection, and service discovery before data transmission can begin. The energy consumed by advertising is estimated to be 219.37 µJ, while service discovery accounts for an additional 576.45 µJ. These results indicate that BLE incurs substantial time and energy overhead before the first packet can be transmitted. In contrast, as shown in Fig. \ref{fig:ble_esb_wakeup_first_packet}(b), ESB behaves markedly differently. After a simpler program initialization, ESB can immediately begin transmitting data packets without any additional connection procedure, thereby avoiding the large pre-transmission overhead observed in BLE.

This difference is further quantified. As shown in Fig. \ref{fig:ble_esb_wakeup_first_packet} (c), BLE requires an average warm-up time of 218.96 ms, whereas ESB requires only 12.51 ms, indicating a substantially shorter transition to the communication state. Correspondingly, Fig. \ref{fig:ble_esb_wakeup_first_packet} (d) shows that BLE consumes 1226.55 µJ during warm-up, while ESB consumes 61.71 µJ. These results consistently demonstrate that ESB achieves approximately one-twentieth the wake-up time and wake-up time energy consumption of BLE, highlighting its superior efficiency for frequent wake-up operations.

These results indicate that, for applications involving frequent wake-up and sleep cycles, ESB can shorten the response time after each wake-up event while also significantly reducing the overall energy consumption.

\subsection{Comparison Between BLE and ESB in Bidirectional Communication}

\begin{figure}[t]
    \centering
    \includegraphics[width= 0.9\linewidth]{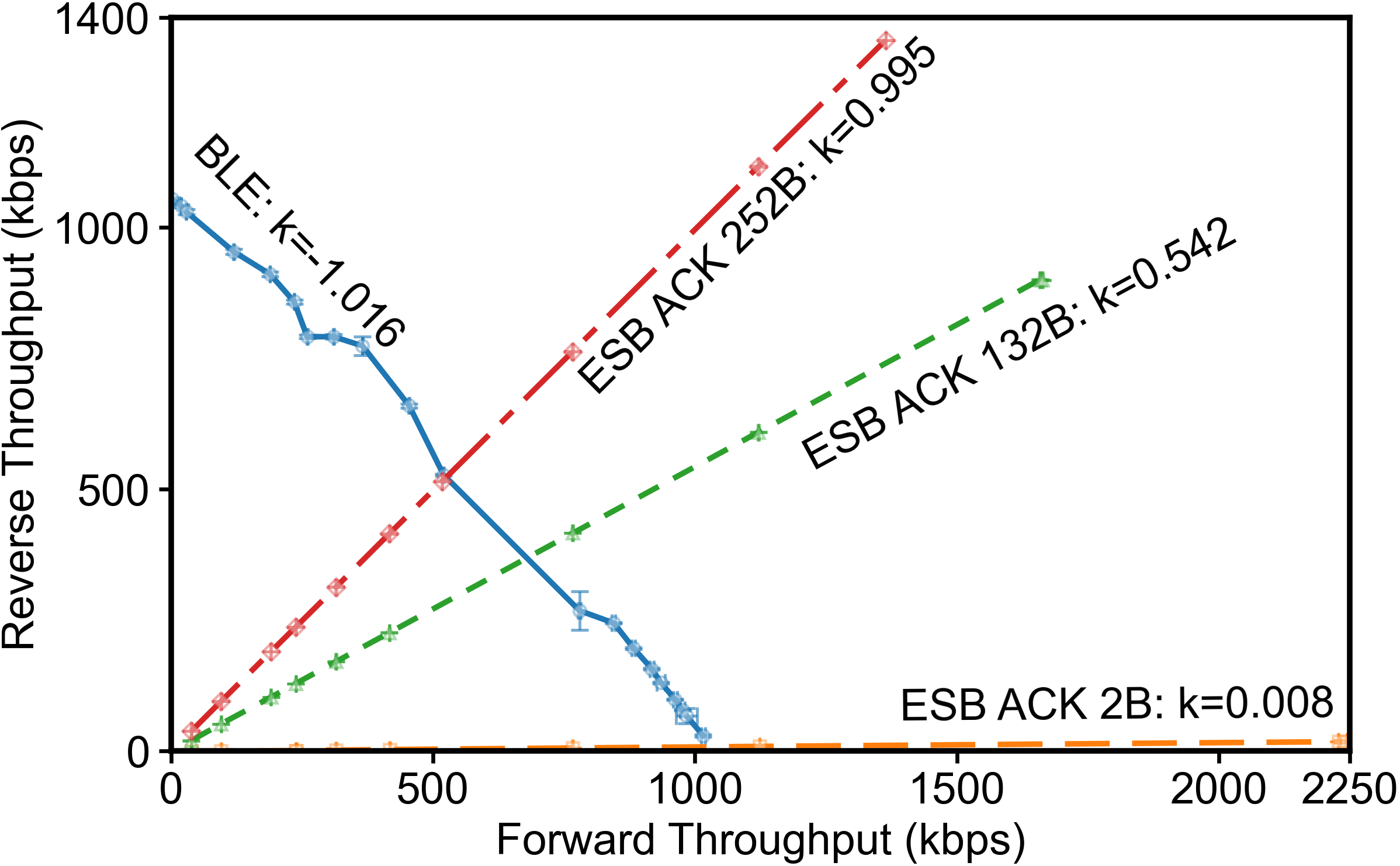}
    \caption{Bidirectional throughput of BLE and ESB. Forward throughput denotes transmission from peripheral to central (or transmitter to receiver), and reverse throughput denotes the opposite direction. k is the fitted slope.}
    \label{fig:bidir}
\end{figure}

Fig. \ref{fig:bidir} compares the bidirectional throughput of BLE and ESB under the controlled-rate methodology. For BLE, the peripheral-to-central and central-to-peripheral throughputs exhibit a strong negative correlation (k=-1.016). As the forward throughput increases from approximately 1 kbps to 1016 kbps, the reverse throughput decreases from about 1054 kbps to 30 kbps. The operating points follow an approximately linear trade-off, while the aggregate throughput remains nearly constant at around 1100 kbps. This indicates that maximizing unidirectional throughput in BLE requires reducing traffic in the opposite direction.

In contrast, ESB demonstrates a fundamentally different behavior governed by ACK payload size. With 2-byte ACK packets, the reverse throughput remains negligible (k=0.008), despite forward throughput reaching up to 2244 kbps, indicating that a minimal ACK payload provides almost no effective reverse data capacity. Increasing the ACK size to 132 bytes significantly enhances reverse throughput (k=0.542). When the ACK size is further increased to 252 bytes, forward and reverse throughput become nearly symmetric (k=0.995), achieving approximately 1364 kbps and 1357 kbps, respectively. These findings indicate that achieving the maximum unidirectional forward throughput in ESB requires minimizing the ACK payload size at the peer node while operating the forward link at saturation.

To quantify this relationship, let $M$ denote the ACK payload size in bytes. The slope and maximum throughputs in each direction are modeled as
\begin{align}
    k(M) &= \frac{M}{252}, \label{eq:k}\\
    F_\mathrm{max}(M) &= \frac{835000}{M + 370} \quad \text{(kbps)}, \label{eq:fmax}\\
    R_\mathrm{max}(M) &= \frac{3408\,M}{M + 370} \quad \text{(kbps)}, \label{eq:rmax}
\end{align}
where $k(M)$ describes the ratio between the reverse ACK-payload throughput and the forward data throughput before saturation, $F_\mathrm{max}(M)$ denotes the maximum achievable forward throughput, and $R_\mathrm{max}(M)$ denotes the maximum achievable reverse throughput carried by ACK payloads. Here, 252 is the maximum forward payload size, and 370 represents the effective per-exchange protocol overhead, including packet headers and radio turnaround time. Fitted to the three saturation measurements ($M \in \{2, 132, 244\}$ B), the model reproduces $F_\mathrm{max}$ within 0.3\% and $k$ within 3.3\%. As $M$ increases, $F_\mathrm{max}$ decays hyperbolically while $R_\mathrm{max}$ rises monotonically, reflecting the inherent trade-off imposed by shared radio airtime.

\begin{figure}[t]
    \centering
    \includegraphics[width=0.9\linewidth]{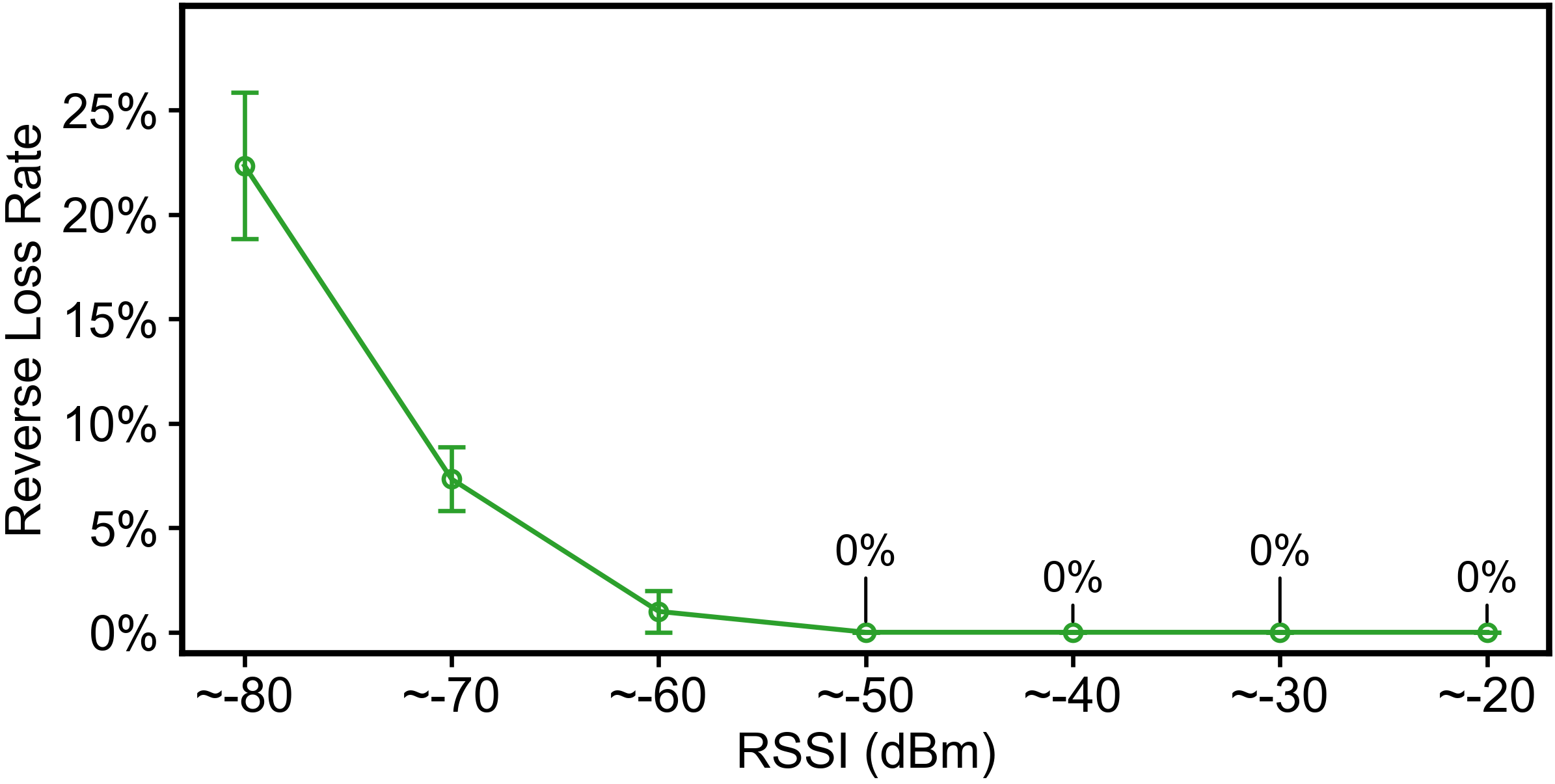}
    \caption{Measured reverse channel packet loss rate of ESB under different RSSI levels.}
    \label{fig:esb_ack_loss_rssi}
\end{figure}

The reverse throughput in ESB may experience data loss, as the ACK packet itself does not support ACK or retransmission mechanisms. Fig. \ref{fig:esb_ack_loss_rssi} shows the estimated reverse-channel packet loss rate as a function of RSSI. At a weak RSSI level of approximately -80 dBm, the ACK back-channel loss rate reaches about 22\%, indicating that ACK-return failures become non-negligible and that reliable reverse-link communication cannot be supported. As the RSSI increases, the loss rate decreases sharply, falling to about 7\% at -70 dBm and around 1\% at -60 dBm. When the RSSI is higher than approximately -50 dBm, no observable loss is recorded within the measurement window. These results indicate that ACK-based reverse transmission in ESB is sensitive and not reliable under weak-link conditions, but can provide reliable reverse-link delivery once sufficient link margin is available.

The results highlight the fundamental limitation of ESB in bidirectional communication. BLE exhibits a shared-capacity trade-off between the two directions, whereas ESB relies on forward-triggered reverse transmission. Consequently, BLE provides symmetric control under constrained shared capacity, while ESB achieves reverse throughput at the cost of link dependency and reliability.

\subsection{Comparison Between BLE and ESB in Security and Ecosystem}

BLE and ESB also exhibit fundamentally different security capabilities. BLE integrates mature security mechanisms, including link-layer AES-128 encryption, Secure Connections based on Elliptic Curve Diffie--Hellman (ECDH) key exchange, and device authentication through pairing procedures \cite{Barua2022Security, Casar2022A}. In contrast, ESB transmits payloads in plaintext, allowing any device configured with the same address and protocol parameters to receive or inject packets without authentication. As a result, encryption, authentication, and replay protection must be implemented at higher layers when ESB is used in security-sensitive applications.

The two protocols further differ in ecosystem interoperability. BLE is an open standard defined by the Bluetooth SIG and is natively supported by smartphones, tablets, and laptops, enabling direct integration with existing consumer and IoT infrastructure. In addition, single-mode BLE device shipments are projected to grow at a compound annual growth rate of 22\% \cite{Zhou2026Reevaluating}, further highlighting its expanding ecosystem. In contrast, although ESB is compatible with other 2.4 GHz wireless technologies at the hardware level, it is currently a proprietary protocol supported only by Nordic Semiconductor radios. Communication requires both endpoints to be preconfigured with identical addresses, PHY settings, and protocol parameters, and no standardized discovery or negotiation mechanism is provided.
\section{Results of BLE-ESB Integration}

\subsection{Enhanced-BLE Coexistence Configuration}
\begin{figure}[t]
    \centering
    \includegraphics[width=0.9\linewidth]{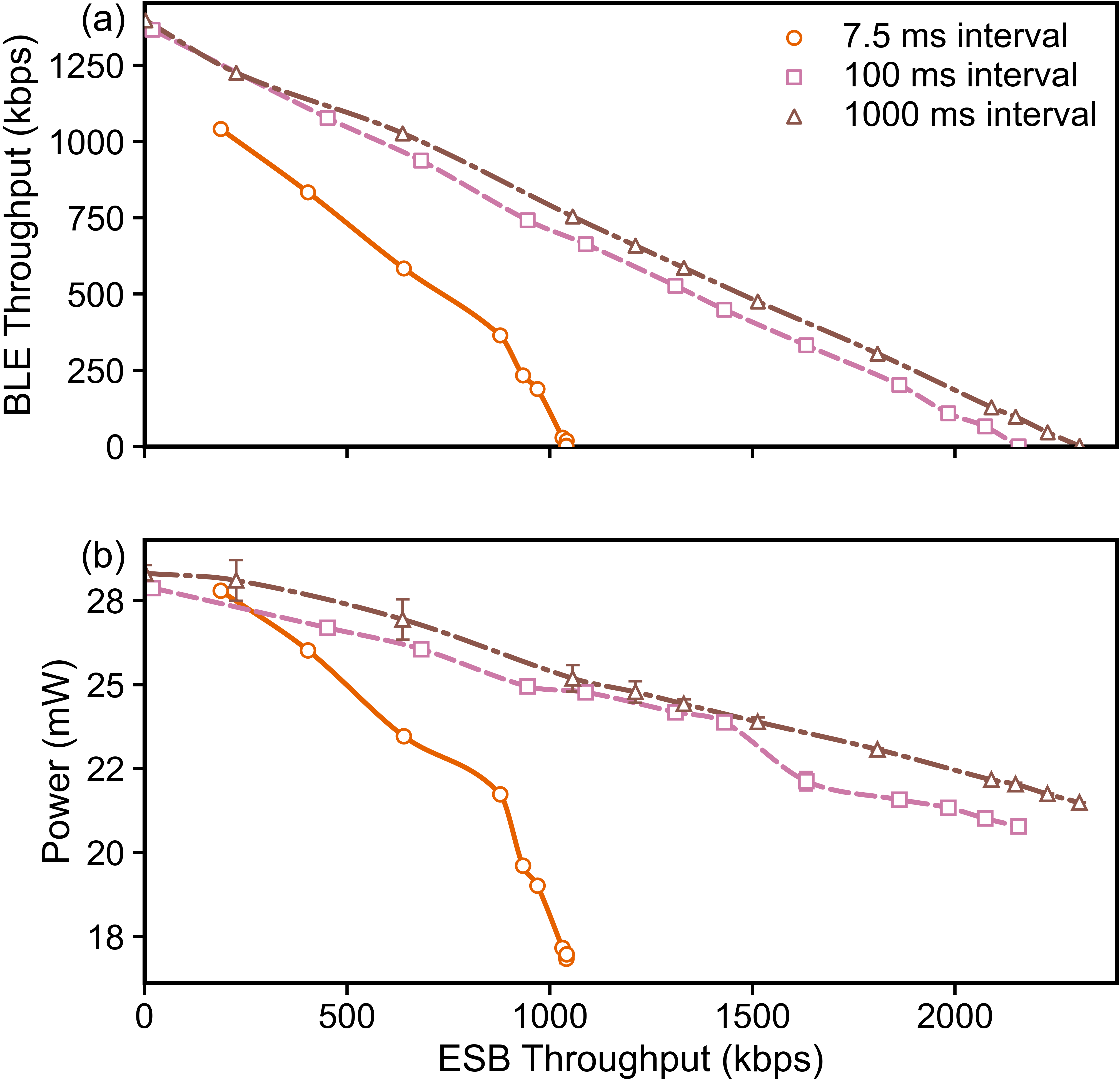}
    \caption{BLE parameter selection for Hybrid Enhanced-BLE coexistence. Larger BLE connection intervals expand the coexistence operating range between BLE and ESB, while higher ESB throughput contribution results in lower overall system power consumption.}
    \label{fig:hybrid_enhanced_ble_coexistence_configuration}
\end{figure}

The coexistence behavior of the Enhanced-BLE framework was evaluated by characterizing the joint throughput and power consumption of the two protocols under varying BLE connection intervals. As shown in Fig. \ref{fig:hybrid_enhanced_ble_coexistence_configuration}, the BLE connection interval serves as the primary control parameter governing the radio resource allocation between BLE and ESB within the MPSL scheduling framework.

Under all three evaluated BLE connection intervals, the BLE and ESB throughput exhibit an inverse relationship due to competition for the shared radio resource, as shown in Fig. \ref{fig:hybrid_enhanced_ble_coexistence_configuration}(a). At the shortest evaluated interval of 7.5 ms, BLE connection events are scheduled more frequently and occupy the shared 2.4 GHz radio more often. Since BLE and ESB cannot use the same radio simultaneously, the short interval leaves fewer idle gaps for ESB transmission. In addition, frequent BLE events introduce extra radio switching and scheduling overhead, reducing the effective data transmission time for BLE itself\cite{Afaneh2022Bluetooth5Throughput}. Therefore, the increased scheduling frequency does not translate into higher useful throughput; instead, both BLE and ESB are constrained to approximately 1000 kbps.

As the BLE connection interval is progressively extended, BLE occupies the shared radio with decreasing frequency, providing ESB with increasingly broader transmission opportunities. Consequently, the achievable ESB throughput increases toward its standalone maximum, reaching approximately 2200 kbps at a 100 ms interval and 2300 kbps at a 1000 ms interval. Meanwhile, the maximum BLE application throughput also increases at larger connection intervals due to improved packet utilization efficiency within each connection event \cite{Afaneh2022Bluetooth5Throughput}.

Fig. \ref{fig:hybrid_enhanced_ble_coexistence_configuration}(b) illustrates the trend of total system power consumption as a function of ESB throughput. Under all three evaluated BLE connection intervals, increasing the ESB throughput contribution leads to reduced overall system power consumption. This behavior originates from the lightweight ESB protocol architecture and its higher transmission efficiency, which improves the overall system energy efficiency. 

To optimize latency and stability, a shorter BLE connection interval is preferred\cite{Thorsrud2020ConnectionIntervalStability, Li2023ApplicationLayer}, whereas a longer connection interval provides a wider coexistence operating range. Therefore, a BLE connection interval of 100 ms was selected as the optimal setting, because it provides a coexistence range comparable to that of 1000 ms while maintaining acceptable latency and stability.

\subsection{Enhanced-BLE Handover Mechanism}

\begin{table}[t]
\centering
\caption{handover latency under different operating scenarios.}
\label{tab:enhanced_ble_handover_latency}
\setlength{\tabcolsep}{5pt}
\renewcommand{\arraystretch}{1.25}
\begin{tabular}{lll}
\hline
Scenario & Case & Latency (ms) \\
\hline
Concurrent operation & BLE adjustment & 35.51 ± 2.81 \\
                     & ESB adjustment & 18.07 ± 2.04 \\
\hline
Standby mode         & BLE-to-ESB & 18.64 ± 2.58 \\
                     & ESB-to-BLE & 49.47 ± 29.16 \\
\hline
Shutdown mode        & BLE-to-ESB & 18.30 ± 1.07 \\
                     & ESB-to-BLE & 309.52 ± 105.09 \\
\hline
\end{tabular}
\end{table}

The measured handover latencies are summarized in Table \ref{tab:enhanced_ble_handover_latency}. When BLE and ESB operate concurrently, changing the ESB allocation shows a faster response than changing the BLE allocation. The latency for ESB adjustment is 18.07 ms, while the latency for BLE adjustment is 35.51 ms. This result indicates that, when both protocols are already active, ESB can respond to bandwidth reallocation more rapidly, whereas BLE requires a longer time before the updated transmission behavior appears in the power waveform.

When only one protocol is transmitting, and the other protocol remains in standby mode, BLE-to-ESB switching can still be completed quickly, with a latency of 18.64 ms. This value is very close to the ESB adjustment latency in the concurrent-operation case. In comparison, ESB-to-BLE switching takes 49.47 ms, which is longer and more unstable. This is mainly because BLE is connection-event-based, so the next effective BLE transmission depends on the timing of the BLE connection event schedule.

When the inactive protocol is completely shut down, BLE-to-ESB switching remains almost unchanged, with a latency of 18.30 ms. This suggests that ESB has a small start-up overhead and can recover rapidly even from the shutdown state. However, ESB-to-BLE switching increases significantly to 309.52 ms. Compared with the standby case, the latency is increased by about 6 times. This large delay is caused by the BLE reinitialization and recovery process, which dominates the total switching time when BLE is powered off.

Overall, the results show that switching toward ESB is consistently fast in all three scenarios, remaining around 18 ms. In contrast, switching toward BLE is strongly affected by whether BLE is kept in standby or completely shut down. Therefore, keeping BLE in standby is more suitable for latency-sensitive applications that require fast switching back to BLE, while shutting down BLE is more suitable for low-power operation when a longer recovery time can be tolerated.

\subsection{Enhanced-BLE Wake-Up Mechanism}
\begin{figure*}[t]
    \centering
    \includegraphics[width=0.9\textwidth]{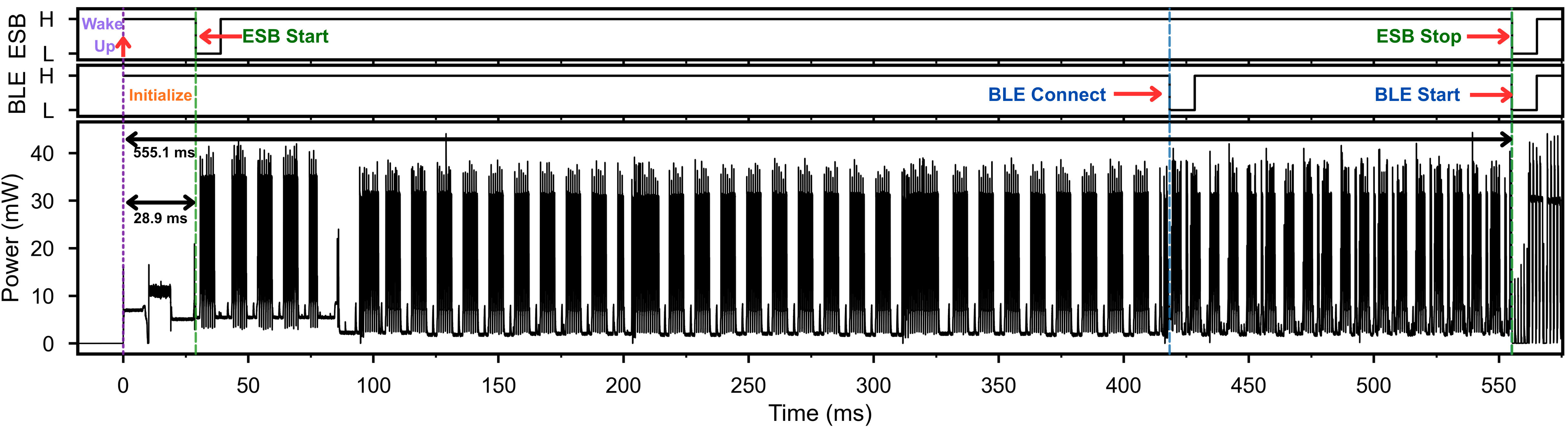}
    \caption{Hybrid Enhanced-BLE wake-up mechanism. ESB is activated immediately after system wake-up to provide rapid communication before BLE connection establishment. After BLE becomes available, ESB is shut down, and the system switches to BLE-dominated communication.}
    \label{fig:hybrid_enhanced_ble_wakeup_mechanism}
\end{figure*}

The hybrid wake-up mechanism was evaluated by recording the system power consumption and protocol state transitions during a complete sleep--wake--transmit--cycle using standby handover. Two GPIO pins were employed as real-time status indicators for BLE and ESB activity, respectively, with both signals synchronized to PPK to enable precise temporal alignment of the power trace with protocol-level events.

As shown in Fig. \ref{fig:hybrid_enhanced_ble_wakeup_mechanism}, the power profile following system wake-up exhibits two distinct phases. Upon wake-up, the ESB transceiver is activated immediately, and the first data packet is transmitted within approximately 28.9 ms, before the BLE connection is established. During this initial ESB phase, BLE advertising and the subsequent connection setup proceed concurrently in the background. Notably, the ESB wake-up time increases by 14 ms compared with the ESB-only configuration shown in Fig. \ref{fig:ble_esb_wakeup_first_packet}, which may be attributed to the additional delay introduced by MPSL and BLE protocol stack initialization.

At the commonly used IoT advertising interval of 100 ms, the BLE connection is established 418.3 ms after wake-up, after which the peripheral starts requesting the service profile. At 555.1 ms after wake-up, the service discovery is completed, ESB transmission is terminated, and data transmission is handed over to BLE to provide a safer or more reliable reverse communication link, as further discussed in the next chapter. The approximately 20 ms delay between BLE start and the observable power change is consistent with the standby handover latency.

This hybrid strategy directly addresses the fundamental limitation of BLE in event-driven IoT applications. Under conventional BLE-only wake-up, the device must complete advertising, scan response, and service discovery before any data can be transmitted. This hybrid approach from ESB to BLE ensures data continuity and integrity throughout the wake-up period.

\subsection{Enhanced-BLE Bidirectional Communication}

\begin{figure}[t]
    \centering
    \includegraphics[width=0.9\linewidth]{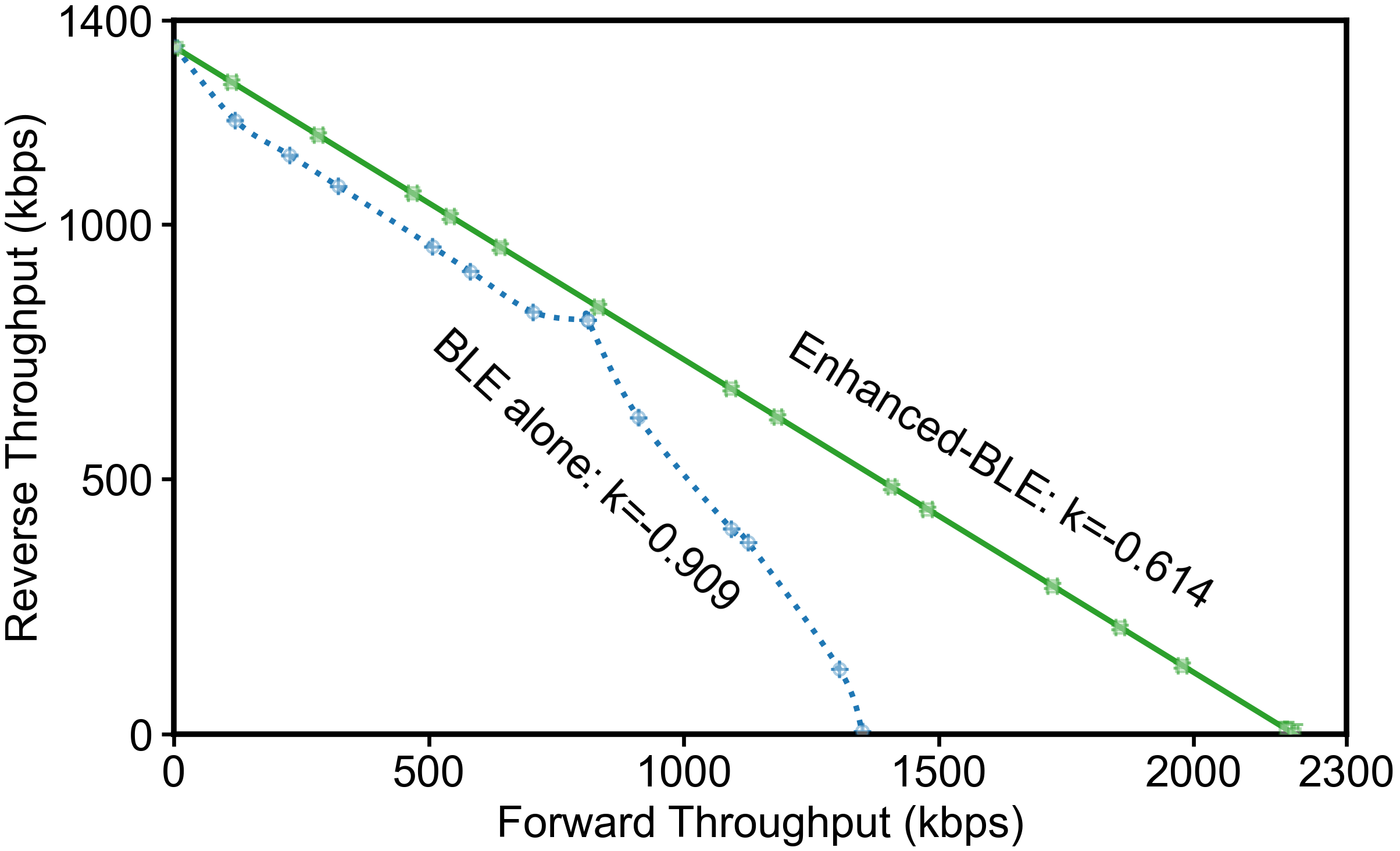}
    \caption{Hybrid Enhanced-BLE bidirectional communication. BLE provides a reverse communication channel for ESB, while ESB supports high-throughput forward transmission for BLE, improving the forward--reverse throughput trade-off compared with single-protocol BLE communication.}
    \label{fig:hybrid_enhanced_ble_bidirectional_communication}
\end{figure}

Fig. \ref{fig:bidir} characterizes the bidirectional communication performance of BLE and ESB, where BLE is configured with a connection interval of 7.5 ms. Since a 100 ms BLE connection interval is selected in the coexistence mode, we further investigated the bidirectional communication performance of a single BLE using a 100 ms connection interval.

The experimental results are shown in Fig. \ref{fig:hybrid_enhanced_ble_bidirectional_communication}. With a 100 ms connection interval, BLE achieves approximately 1350 kbps in both the forward and reverse directions. The fitted slope is k=-0.91, which is close to -1 and generally consistent with the trend obtained under the 7.5 ms interval. In contrast, Enhanced-BLE achieves a maximum forward throughput of 2200 kbps by leveraging the high forward bandwidth of ESB, while also reaching a maximum reverse throughput of 1350 kbps through reliable BLE-based feedback transmission. The Enhanced-BLE curve exhibits a flatter trend, with a fitted slope of k=-0.61.

This architecture addresses the bandwidth limitation of BLE, enabling low-power 2.4 GHz communication to achieve an almost two-fold improvement in forward bandwidth compared with BLE. As a result, the operating range of the overall system is expanded, and any throughput requirement below the curve can be supported. Meanwhile, the architecture avoids the instability and packet loss associated with ESB-based reverse transmission, thereby enabling stable and efficient bidirectional communication.

\subsection{Enhanced-BLE TXP and PHY Control}
\begin{figure}[t]
    \centering
    \includegraphics[width=0.9\linewidth]{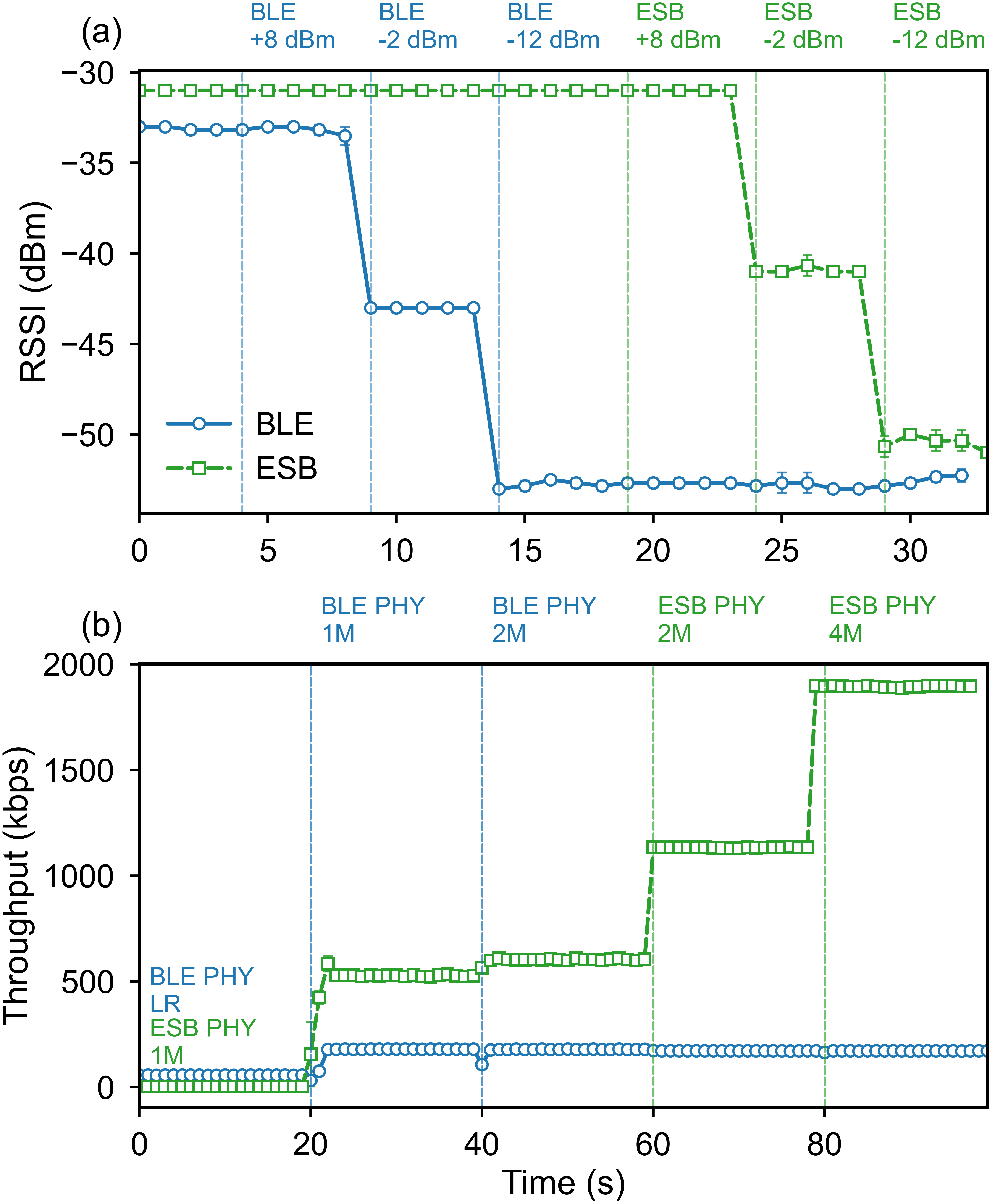}
    \caption{Hybrid Enhanced-BLE TXP and PHY control. (a) RSSI response when BLE and ESB TXP levels are independently adjusted. (b) Throughput response when BLE and ESB PHY configurations are separately controlled, demonstrating protocol-level flexibility in Hybrid Enhanced-BLE communication.}
    \label{fig:hybrid_enhanced_ble_txp_phy_control}
\end{figure}

Fig. \ref{fig:hybrid_enhanced_ble_txp_phy_control} (a) validates independent TXP control. As BLE TXP is stepped from +8 to -2 to -12 dBm, BLE RSSI decreases in roughly 10 dB steps from approximately -33 dBm to -43 dBm, while ESB RSSI remains stable at approximately -31 dBm. The subsequent ESB sweep produces symmetric behavior: ESB RSSI changes accordingly while BLE RSSI remains unchanged. The orthogonality of the two RSSI traces confirms that BLE and ESB TXP control are effectively decoupled at the radio, since each protocol independently writes and reads the TXP register within its own MPSL-allocated radio timeslot. These results further indicate that existing advanced TXP control algorithms can be independently applied to each protocol \cite{11222604,10.1145/2746342,PARK2020107520}, enabling system-level power optimization without cross-protocol interference.

Fig. \ref{fig:hybrid_enhanced_ble_txp_phy_control}(b) validates the separate PHY configuration. After BLE switches from the long-range coded PHY (S8) to the 1M PHY, its throughput increases from approximately 50 kbps to 200 kbps after stabilization. Meanwhile, the steady-state ESB throughput increases from 0 kbps to approximately 500 kbps. This improvement can be attributed to the higher PHY efficiency, which increases the BLE transmission throughput and releases more radio time slots for ESB transmission. During PHY switching, transient jumps are observed in both BLE and ESB throughput because the radio occupancy of BLE changes dynamically during the PHY update process. When BLE switches from the 1M PHY to the 2M PHY, the BLE throughput remains nearly unchanged, as the current packet generation rate can be fully supported by both PHY configurations. In contrast, the ESB throughput exhibits a slight increase under the 2M PHY, since the higher PHY efficiency provides additional idle intervals for ESB transmission. These results indicate that the BLE PHY configuration not only affects BLE throughput but also influences ESB performance through radio-time allocation.

When the ESB PHY is increased to 2M, its throughput correspondingly increases to approximately 1200 kbps. After the ESB PHY is further increased to 4M, the throughput approaches 1800 kbps. Throughout these transitions, BLE throughput remains stable at around 200 kbps, indicating that ESB PHY reconfiguration does not affect BLE performance. This behavior reflects the design of the framework, where BLE is prioritized, and ESB operates opportunistically within the remaining scheduled radio resources. Advanced PHY control algorithms can be applied to Enhanced-BLE for protocol-specific performance optimization \cite{Spork2020Improving, Sheikh2021Adaptive}.

Together, these results establish that the Enhanced-BLE framework supports per-protocol adaptive radio management: BLE and ESB TXP can be tuned to different values on the same hardware, and each protocol's PHY can be separately selected according to its throughput or range requirements.

\section{Conclusion and Future Work}

This work systematically benchmarked BLE and ESB on a unified Nordic nRF54L15 platform and demonstrated that the two protocols exhibit fundamentally complementary communication characteristics. Compared with BLE, ESB significantly reduces single-packet transmission time and energy consumption, doubles the achievable maximum throughput, and greatly lowers wake-up latency and energy overhead due to its lightweight connectionless architecture and 4M PHY capability. However, ESB also exhibits several important limitations compared to BLE, including unreliable reverse transmission caused by the lack of ACK and retransmission support, as well as the absence of native interoperability and security mechanisms.

Motivated by these complementary properties, we proposed Enhanced-BLE, a hybrid 2.4 GHz communication framework based on MPSL-enabled BLE--ESB coexistence. Experimental results show that the BLE connection interval governs radio resource allocation between the two protocols, with a 100 ms interval providing optimal balance between stability and operating range. The handover results demonstrate that the proposed framework can dynamically reconfigure protocol allocation with low transition latency across different operating scenarios. The ESB-assisted wake-up mechanism enables packet transmission before BLE service discovery is completed, thereby reducing the communication gap. In bidirectional communication, the Enhanced-BLE strategy expands the achievable throughput region by combining high ESB forward throughput with reliable BLE reverse transmission. Furthermore, TXP can be independently configured for BLE and ESB with negligible cross-protocol influence, whereas BLE PHY selection affects ESB throughput through radio-time occupancy under MPSL scheduling, while ESB PHY reconfiguration introduces negligible impact on BLE performance.

Future work will extend Enhanced-BLE beyond platform-specific implementation toward broader multi-vendor and multi-protocol deployment. First, the ESB and MPSL code will be ported to chipsets from additional vendors to evaluate the portability of the hybrid BLE--ESB framework. This migration is expected to be feasible because ESB, MPSL, and BLE share compatibility with 2.4 GHz radio hardware, which may reduce the implementation barrier beyond Nordic platforms. Second, MPSL coexistence with other 2.4 GHz protocols, including Zigbee, Thread, and Matter, will be evaluated to investigate its effectiveness in practical IoT environments. Third, adaptive TXP and PHY control algorithms will be developed to jointly optimize BLE and ESB operation according to link quality, traffic demand, and energy constraints.

\bibliographystyle{IEEEtran}
\bibliography{main}

 \vspace{-30pt}


\begin{IEEEbiography}[{\includegraphics[width=1in,height=1.25in, clip,keepaspectratio]{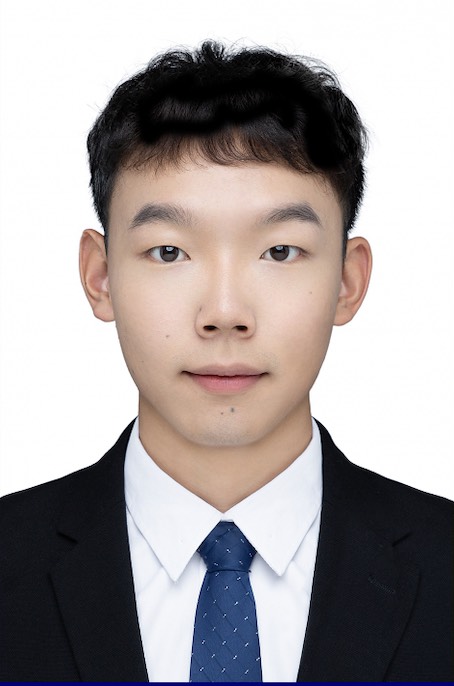}}]{Ziyao Zhou} received the B.Eng. degree in Communication Engineering from the University of Electronic Science and Technology of China, Chengdu, China, in 2023, and the M.Sc. degree in Communication Engineering from Nanyang Technological University, Singapore, in 2024. He is currently working toward a Ph.D. in the Digital, AI, Robotics, and Electronics (DARE) Lab for Translational Medicine, School of Electrical and Electronic Engineering, Nanyang Technological University. His research interests include wireless sensor and actuator networks for gastrointestinal tract applications.
\end{IEEEbiography}

 \vspace{-10pt}

\begin{IEEEbiography}[ {\includegraphics[width=1in,height=1.25in,clip,keepaspectratio]
{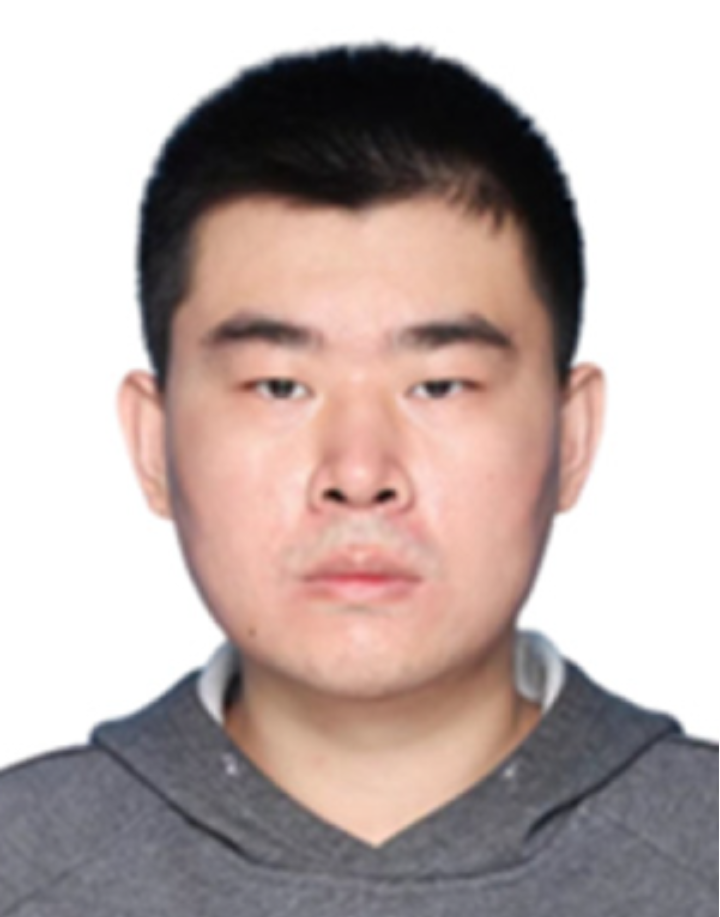}}
]{Chen Shen} received the B.Eng. degree in electrical and electronic engineering from the University of Nottingham, Ningbo, China, in 2014, and the M.Sc. and
Ph.D. degrees in electrical and electronic engineering from Nanyang Technological University, Singapore, in 2019 and 2025, respectively.
He is currently a Research Fellow in the Digital, AI, Robotics, and Electronics (DARE) Lab for Translational Medicine, School of Electrical and Electronic Engineering, Nanyang Technological University.  His research interests
include low-power ASIC design, neuromorphic computing, edge AI, and ingestible electronics.
\end{IEEEbiography}
 \vspace{-10pt}

\begin{IEEEbiography}[{\includegraphics[width=1in,height=1.25in,clip,keepaspectratio]{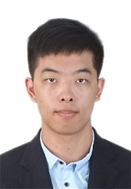}}]{Tiancheng Cao} received the B.Eng. (Hons.) degree and the Ph.D. degree from Nanyang Technological University (NTU), Singapore, in 2021 and 2024, respectively. He was supported by the NTU Science and Engineering Undergraduate Scholarship and the Nanyang President Graduate Scholarship, and received the Best Ph.D. Thesis Award. He is currently a Schmidt AI in Science Fellow at the Centre for System Intelligence and Efficiency (CSIE), NTU, supported by Schmidt Sciences. His research focuses on neuromorphic computing, edge AI, biomedical systems, Internet of Medical Things (IoMT), and translational medicine.
\end{IEEEbiography}
 \vspace{-10pt}
\begin{IEEEbiography}[{\includegraphics[width=1in,height=1.25in, clip,keepaspectratio]{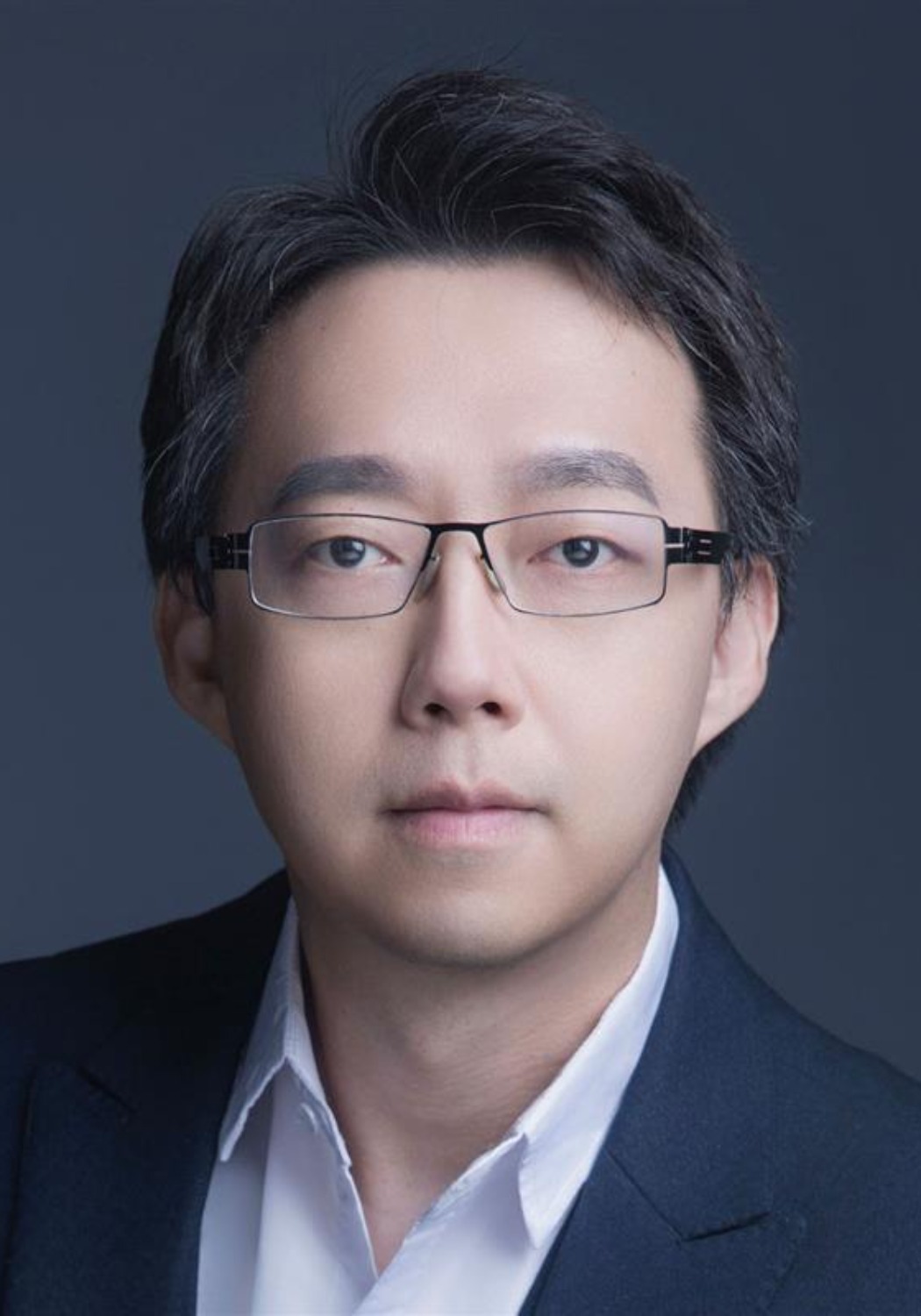}}]{Hen-Wei Huang} received the B.S. and M.S. degrees in Mechanical Engineering from National Taiwan University, Taiwan, in 2011 and 2012, respectively, and the Ph.D. degree in Robotics from ETH Zürich, Switzerland, in 2018.
He was a faculty member in Medicine at Harvard Medical School and an Associate Scientist at Brigham and Women’s Hospital, Boston, USA, from 2021 to 2024, where he contributed to the development of advanced biomedical sensing and robotic systems.
He previously worked as an R\&D Engineer at TSMC (2012–2013) and MedSense Inc. (2013–2014), focusing on advanced sensing and electronic packaging technologies.
He is currently an Assistant Professor at the School of Electrical and Electronic Engineering, with a joint appointment at the Lee Kong Chian School of Medicine, Nanyang Technological University (NTU), Singapore, since 2024. He is also the Director of the DARE (Digital, AI, Robotics, and Engineering) Laboratory for Translational Medicine.
His research focuses on in vivo wireless sensor networks, ingestible and implantable electronics, robotic-assisted drug delivery, and translational medical technologies, with the goal of enabling next-generation minimally invasive diagnostics and therapeutics.
\end{IEEEbiography}



\end{document}